# Comparison and characterization of efficient frequency doubling at 397.5 nm with PPKTP, LBO and BiBO crystals

**Xin WEN**[1, 2], **Yashuai HAN**[1, 2], and **Junmin WANG**[1, 2, 3, *]

[1] State Key Laboratory of Quantum Optics and Quantum Optics Devices (Shanxi University),
[2] Institute of Opto-Electronics, Shanxi University,
[3] Collaborative Innovation Center of Extreme Optics (Shanxi University),
No.92 Wu Cheng Road, Tai Yuan 030006, Shan Xi Province, People's Republic of China

E-mail:* wwjjmm@sxu.edu.cn

**Abstract:** A continuous-wave Ti:sapphire laser at 795 nm is frequency doubled in a bow-tie type enhancement four-mirror ring cavity with $LiB_3O_5$ (LBO), $BiB_3O_6$ (BiBO), and periodically polled $KTiOPO_4$ (PPKTP) crystals, respectively. The properties of 397.5 nm ultra-violet (UV) output power, beam quality, stability for these different nonlinear crystals are investigated and compared. For PPKTP crystal, the highest doubling efficiency of 58.1% is achieved from 191 mW of 795 nm mode-matched fundamental power to 111 mW of 397.5 nm UV output. For LBO crystal, with 1.34 W of mode-matched 795 nm power, 770 mW of 397.5 nm UV output is achieved, implying a doubling efficiency of 57.4%. For BiBO crystal, with 323 mW of mode-matched 795 nm power, 116 mW of 397.5 nm UV output is achieved, leading to a doubling efficiency of 35.9%. The generated UV radiation has potential applications in the fields of quantum physics.
**Keywords:** frequency doubling, enhancement cavity, UV laser, PPKTP crystal, LBO crystal;
BiBO crystal
**PACS:** 42.60.Da, 42.70.Mp, 42.72.Bj, 42.79.Nv

## 1. Introduction

Ultra-violet (UV) laser is in great demand for many practical applications, such as bio-medicine, laser printing, high-resolution spectroscopy, and quantum optics. The UV laser at 397.5 nm can be used to pump an optical parametric oscillator (OPO), for the generation of squeezed or entangled states at 795 nm. The generated lights are resonant on rubidium (Rb) $D_1$ line, and can be useful in the fields of precise measurements [1], light-atom interaction [2, 3], quantum memory [4], and quantum networks [5].

Second harmonic (SH) generation is employed for obtaining the 397.5 nm UV radiation. The heavy absorption at these UV wavelengths and the associated thermal effect, however, will limit the frequency doubling. Therefore, it is crucial to choose a proper nonlinear crystal. Besides the effective nonlinear coefficient that determines the conversion efficiency, the transparent range that relates to the absorption and the stability which decides the future utility are also necessary to be considered.

Experimental generations of the UV radiation by frequency doubling have been demonstrated in recent years. Periodically polled $KTiOPO_4$ (PPKTP) crystal is widely utilized [6-9]. Our group frequency doubled the 795 nm laser with PPKTP crystals in a bow-tie type four-mirror ring cavity [10] and a semi-monolithic cavity [11], the corresponding conversion efficiency is 31% and 41% respectively in the low power regime. Wang *et al*. demonstrated the frequency doubling at 780 nm with a PPKTP crystal, a maximum conversion efficiency of 12% is obtained [12]. A monolithic PPKTP cavity is used in frequency doubling at 852 nm, achieving a conversion efficiency of 45% [13]. Pizzocaro *et al*. presented an excellent result using the $LiB_3O_5$ (LBO) crystal for doubling at 399 nm. With 1.3 W of 798 nm fundamental power, 1.0 W of 399 nm UV radiation is obtained, the frequency doubling efficiency is 80% [14]. Ruseva and Hald used a $BiB_3O_6$ (BiBO) crystal for frequency doubling to get 384 nm UV radiation, they achieved a 50% of conversion efficiency at just 130 mW coupled fundamental power [15]. β-BaB$_2$O$_4$ (BBO) is used for the frequency doubling at 806 nm, getting a conversion efficiency of 24% [16]. Besides, when the wavelength came into the deep UV regime, KBe$_2$BO$_3$F$_2$ (KBBF) crystal could be more capable to be used [17]. Other crystals, for instance, KNbO$_3$ [18, 19] and LiNbO$_3$ [20], however, cannot be used in the UV regime, for their low limit wavelength is around 400 nm.

Cavities are widely used in frequency conversion process, benefiting from the enhancement effect of the fundamental wave. The works listed above are all the SH generation with external cavities, in which the linear loss and the coupling efficiency from both mode matching and impendence matching of the cavities should be taken into consideration. It has the benefit of easy to maintain the single-mode operation inside the cavity. Intracavity frequency doubling, however, is another choice for frequency conversion. It has the advantages of requiring less pump power, compact in structure and is easy to align and insert optical elements, but extra parts should be added to ensure a single-mode output. An intracavity frequency doubled 397.5 nm UV radiation is obtained with LBO and BiBO crystals in a Ti:Sapphire laser, and a maximum UV power of 1.58W is accomplished [21].

In this paper, we demonstrate the frequency doubling from 795 nm to 397.5 nm in an external enhancement bow-tie type cavity with PPKTP, LBO, and BiBO crystals, respectively. For the purpose of preparing the pump light of the OPO process, the mostly concerned characters of the SH output are investigated here, for instance, the SH output power, beam quality and the long-term stability. PPKTP crystal has the highest conversion efficiency but with the problem of severe thermal effect. LBO crystal has advantages at high power level with great long-term stability, after shaping the elliptical output beam, it could be a wonderful candidate for pumping an OPO. For BiBO crystal, the photo-refractive effect restrains the further increase in efficiency, a well-grown crystal could get better results. This is the first comparison work based on frequency doubling with different nonlinear crystals at 795 nm. Our work provides a probable reference for the choices of nonlinear crystals at these near-infrared wavelengths, and could be helpful to the nonlinear frequency conversion.

## 2. Choices of nonlinear crystals

In the UV regime, for the heavy absorption, the choices are limited. In our previous work [10, 11] PPKTP crystal is utilized. The advanced technique of quasi-phase matching makes it possible to use the largest effective nonlinear coefficient (10.8 pm/V) under a proper poling period and temperature. The limit is the transparent range in our case. The 397.5 nm SH wavelength is close to the edge of the transmission window of KTP [22], the absorption is severe, and problems follow, for instance, the gray tracking and the thermal instability. As a result, high power and long-term working will be the challenge.

Table 1. Parameters of the crystals for frequency doubling from 795 nm to 397.5 nm.

| Parameters | PPKTP | LBO | BiBO | BBO |
|---|---|---|---|---|
| Phase matching | quasi-phase matching Type-I | critical phase matching Type-I (o+o-e) | critical phase matching Type-I (e+e-o) | critical phase matching Type-I (o+o-e) |
| Cutting angle |  | θ=90° φ=32.1° | θ=150.7° φ=90° | θ=29.4° φ=0° |
| d$_{eff}$/pm/V | 10.8 [23] | 0.75 [14] | 3.61 [24] | 2.0 [16] |
| Temperature | 53 ℃ | 25 ℃ | 25 ℃ | 25 ℃ |
| Walk-off angle/mrad | 0 | 16.2 [14] | 40.7 [25] | 6.7 [16] |
| Refractive index | 1.8 [26] | 1.6 [27] | 1.8 [28] | 1.6 [29] |
| Transparent range/nm | 350-4400 [11] | 160-2600 [14] | 280-2500 [30] | 190-3500 |

LBO crystal is popular in frequency doubling, especially in the UV regime. Its wide transparent range, small walk-off, and good long-term stability are all the properties required in the integrated frequency doubler. Although the effective nonlinear coefficient is low, a careful control of intracavity loss, an optimized transmissivity for impedance matching and a high fundamental power will definitely lead to the effective conversion.

BiBO crystal is a relatively newly developed nonlinear material. It can also be used for frequency doubling in UV range with a transparent range of 280-2500 nm. The effective nonlinear coefficient is higher than LBO, but with a larger walk-off angle. As with the LBO crystal, such cost-efficient crystals have great potential in frequency conversion. The main parameters of the three crystals are listed in table 1.

## 3. Experimental setup

The schematic diagram of the frequency doubling system is shown in figure 1. The fundamental source is a continuous-wave Ti:Sapphire laser, whose wavelength is tuned to 795 nm, resonant on the rubidium $D_1$ line, and has the linewidth of less than 50kHz. A 30-dB isolator is used to avoid optical feedback. The phase-type electro-optical modulator introduces a modulation signal to the fundamental source with a frequency of 4.1 MHz, which is used to lock the frequency doubling cavity with the Pound-Drever-Hall side-band method [31]. The half-wave plate before the cavity ensures the polarization required for the phase-matching condition. The bow-tie type frequency doubling cavity consists of two plane mirrors and two concave mirrors with a radius of curvature of 100 mm. M1 serves as the input coupler with an optimized transmissivity for individual crystals (10% for PPKTP, 1.2% for LBO and 3% for BiBO). The other three mirrors are all coated highly reflected ($R_2$, $R_3$ and $R_4 >$ 99.9%) at the fundamental wavelength (FW), and M4 is also highly transmitted ($T_4 > 95\%$) at the SH wavelength. The total cavity length is 600 mm, and the concave mirrors are separated for 120 mm with the nonlinear crystal placed in the center. Such cavity configuration leads to a beam waist of 40 μm inside the crystal.

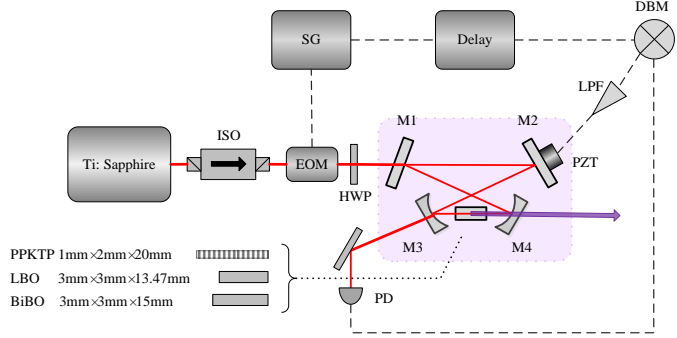

**Figure 1** The schematic diagram of the frequency doubling system. ISO: optical isolator; EOM: phase-type electro-optical modulator; HWP: half-wave plate; PZT: piezoelectric transducer; PD: photodiode; SG: signal generator; DBM: double balanced mixer; LPF: low-pass filter. Solid lines are the optical parts, and the dash lines are the electric parts used to lock the cavity.

PPKTP crystal (Raicol Crystals Ltd.) is 1 mm×2 mm×20 mm with a poling period of 3.15 μm. It is quasi-phase matched with a phase matching temperature of around 53 ℃. The crystal surfaces are anti-reflected coated for both FW and SH wavelength. The crystal is placed in a copper-made oven, whose temperature is precisely controlled by a Peltier element and a temperature controller.

Both LBO and BiBO crystals (Castech Inc.) have a section of 3 mm×3 mm and a length of 13.47 mm for LBO and 15 mm for BiBO, respectively. The crystals are cut normal incidence for critically type-I phase matching, the cutting angle for LBO is θ = 90°, φ = 32.1°, and for BiBO is θ = 150.7°, φ = 90°. The crystals work at room temperature which is not actively stabilized.

The frequency doubling cavity is primarily designed for the PPKTP crystal. In the theory presented by Boyd and Kleinman [32], the optimal beam waist is given with the focus parameter ξ equals to 2.84, it is 22 μm in the 20 mm-long PPKTP crystal. In experiments, however, we use a loose focus as mentioned in references [10, 11] and [33] to compensate the thermal effect. The designed beam waist inside the crystal is 40 μm, nearly twice of the calculated results, such optimization will greatly alleviate the thermal effect and still keep a high conversion efficiency.

Beside, as shown in the theory, SH power will stay near the conversion peaks when the focus parameter ξ changes in a relatively wide range. In our experiments, for LBO crystal, ρ=0.016, B=3.35, ξ=0.66; and for BiBO crystal, ρ=0.041, B=9.40, ξ=0.66, ρ is the double refraction angle and B partially determines the conversion efficiency. Both ξ are within the values which decrease 10% from the maximum point. Therefore, although the cavity configurations are not specifically optimized for LBO and BiBO crystal, the focus conditions are both close to the optimal one, which has few influence in cavity enhanced frequency doubling.

## 4. Experimental results and discussion

The frequency doubling efficiency can be written in equation (1) [33, 34]:

$$\frac{P_c}{P_1} = \frac{T}{(1-\sqrt{(1-T)(1-L)(1-E_{NL}P_c)})^2} \quad (1)$$

Where $P_1$ is the incident fundamental power, $P_c$ is the circulating power, $T$ is the transmissivity of the input coupler, $L$ is the intracavity loss and $E_{NL}$ is the single-pass conversion coefficient. For PPKTP crystal, however, an extra term $\Gamma$ has to be added. As mentioned above, the generated SH at 397.5 nm is at the edge of the transparent window of the PPKTP crystal. The SH absorption loss $\Gamma_{abs}$ cannot be neglected here, $P_{abs}=\Gamma_{abs}P_c^2$. Therefore, the overall nonlinear loss $\Gamma$ should be written as $\Gamma=E_{NL}+\Gamma_{abs}$. For PPKTP, the measured absorption coefficient is around 18%/cm, which is close to the parameters mentioned before [35, 36]. Then the conversion efficiency $\eta=P_2/P_1$ follows the equation in Ref. [33, 37]:

$$\sqrt{\eta} = \frac{4T\sqrt{E_{NL}P_1}}{[2-\sqrt{1-T}(2-L-\Gamma\sqrt{\frac{\eta P_1}{E_{NL}}})]^2} \quad (2)$$

Figure 2-4 show the frequency doubling results for the individual crystals: 20 mm-long PPKTP, 13.47 mm-long LBO and 15 mm-long BiBO. In figure 2, blue squares are the data for the generated SH power, red circles are for the corresponding conversion efficiency, and the lines are the calculated results with the parameters measured in our case: $T = 10\%$, $E_{NL} = 2.3\%/W$, and $L = 1.5\%$. Each point is measured with careful optimization of the temperature. With a maximum fundamental power of 240 mW, we achieved 137 mW of the SH power. The highest conversion efficiency is 58.1% when the mode-matched fundamental power reaches 191 mW, the corresponding SH power is 111 mW. Such result is greatly improved compared with our previous work [10, 11], owing to the reduced intracavity loss. The measured data shows a gradual deviation with the increasing input power. It is mainly the thermal induced bistability that prevents the ideal locking of the cavity at the top of the fringes.

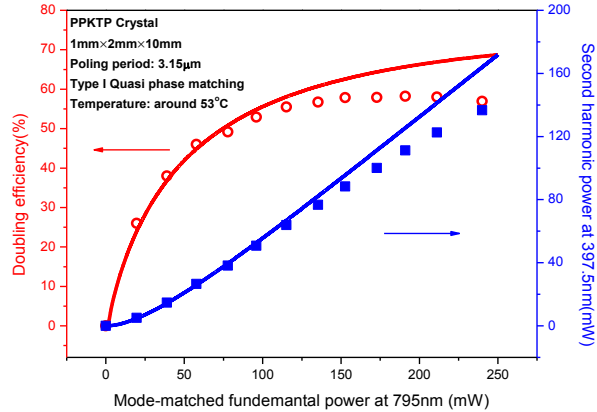

**Figure 2** Frequency doubling results for the 20 mm-long PPKTP crystal. Blue squares are the measured SH power at 397.5 nm, red circles are the corresponding doubling efficiency, and lines are the calculated result.

As shown in figure 3, for LBO crystal, the theoretical curves match well with the experimental measured data. With a mode-matched fundamental power of 1.34 W, we obtain an output SH power of 770 mW, implying a conversion efficiency of 57.4%. For the low efficient nonlinear coefficient, the input coupler should be replaced by a new one with a relatively low transmissivity. With an intracavity loss of 0.49%, $E_{NL}$ of around $6.5 \times 10^{-5}$/W, we calculated the optimum transmissivity to be 1.2%. At current input power, we do not see the saturation, from which we can infer a further growing in doubling efficiency with higher fundamental power.

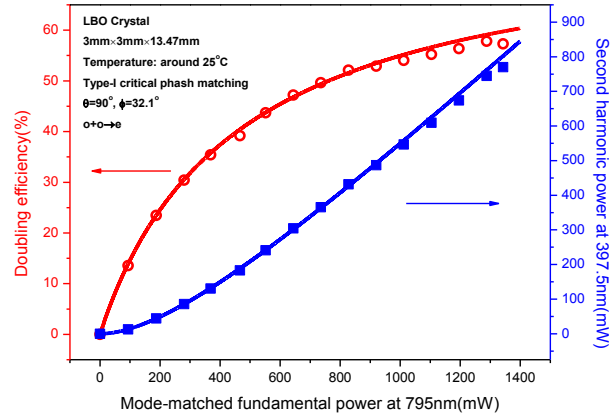

**Figure 3** Frequency doubling results for the 13.47 mm-long LBO crystal. Blue squares are the measured SH power at 397.5 nm, red circles are the corresponding doubling efficiency, and lines are the calculated result.

The experimental results of the BiBO is presented in figure 4. When the mode-matched fundamental power is 323 mW, the obtained SH power is 116 mW, leading to a conversion efficiency of 35.9%. The parameters for calculation is $T = 3\%$, $L = 1\%$, and $E_{NL} = 5.6 \times 10^{-4}$/W. Experiment in the higher power regime is not continued, for we find the instability of the SH power. It cannot be maintained at the initial point, but dithers in a wide range. Such phenomenon has been mentioned in the previous experiments [15, 24], which suggest a photorefractive effect that might be compensated by adjusting the polarization of the FW. Not many experiments have encountered such problem, so we tried another BiBO crystal with a length of 10 mm, but the same problem occurred again. We think it is the diversity from different crystals that result in the instability. Even with the bad behavior in our case, we can still see the potential of the BiBO crystal used in frequency doubling process if it is well produced.

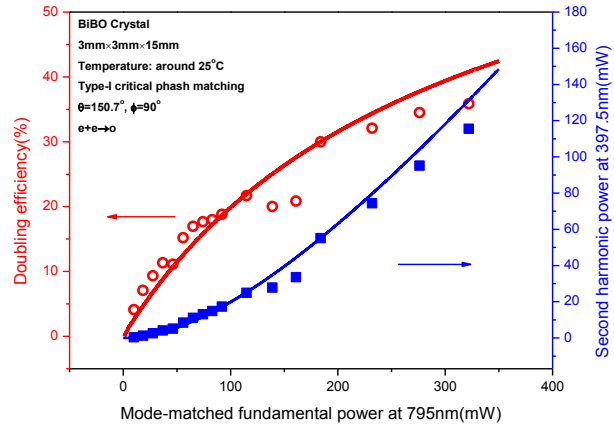

**Figure 4** Frequency doubling results for the 15 mm-long BiBO crystal. Blue squares are the measured SH power at 397.5 nm, red circles are the corresponding doubling efficiency, and lines are the calculated results.

In order to serve as the pump source of the OPO cavity, the long-term stability of the UV radiation is crucial. We evaluate the performance of the PPKTP and LBO crystal respectively. In 30 minutes, the root-mean-square (RMS) fluctuation for PPKTP is 3.5% with a UV power of around 130 mW. For LBO, the absorption of the UV radiation is not observed, regardless of the thermal induced instability, the LBO cavity get a better result. With a UV power of about 530 mW, the 0.5% RMS fluctuation is far better than that of PPKTP. The results are presented in figure 5.

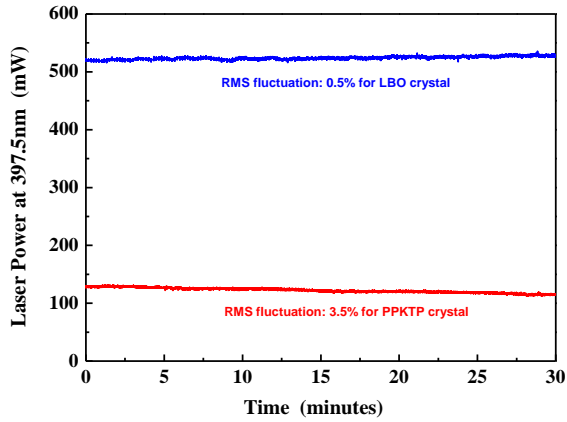

**Figure 5** Power stability of PPKTP and LBO crystals in 30 min. The RMS fluctuation of PPKTP is 3.5% shown in red line, and it is 0.5% for LBO shown in blue line.

Another important issue to be concerned is the beam quality of the generated UV radiation. For PPKTP, the measured $M^2$ are $M_x^2 = 1.19$ and $M_y^2 = 1.16$, as reported in our previous work [10]. Figure 6(a) shows the typical intensity distribution of the output beam from the PPKTP cavity, indicating a good beam quality. The angular phase-matched LBO crystal, however, suffers from the walk-off effect, so the output beam is elliptical in shape. The intensity distribution of the direct UV output is shown in figure 6(b). It is necessary for an anamorphic prism pair to be used, after which the measured $M^2$ in horizontal and vertical directions are $M_x^2 = 1.15$ and $M_y^2 = 1.03$ respectively. The results are presented in figure 6(c), and the inset is the beam intensity distribution after the anamorphic prism pair. The shaped beam has good special mode, and can be easily mode matched in the following OPO cavity. For BiBO crystal, the walk-off angle is larger than LBO crystal, and the asymmetry of the output beam shape is more obvious.

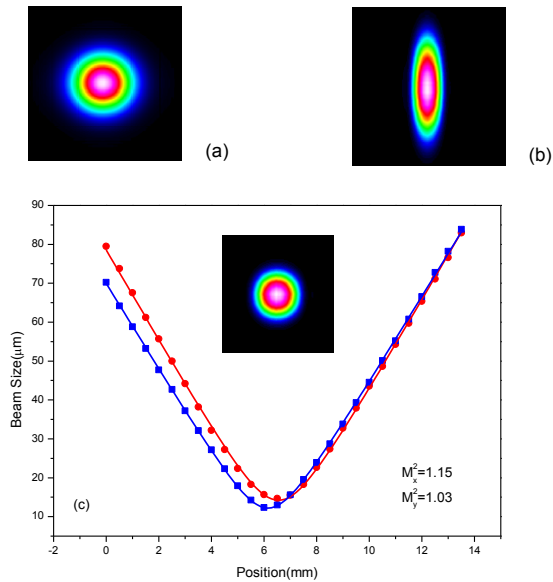

**Figure 6** Typical intensity distribution of the output UV beam and the measured $M^2$ of the LBO cavity. (a). Typical intensity distribution of the PPKTP cavity output beam; (b). Typical intensity distribution of the LBO cavity output beam, elliptical in shape; (c). Measured $M^2$ of the LBO cavity output after shaping, the inset is the intensity distribution after the anamorphic prism pair.

For choosing the optimal crystal, both the fundamental power and the usage should be considered. In low and medium power regime, despite the absorption inside the PPKTP crystal, the thermal effect does not really matter. Therefore, the advantages of high effective nonlinear coefficient and the good SH beam quality make it suitable for the employment of PPKTP crystal, especially for our further aim of pumping the OPO. In high fundamental power regime, instead, LBO will be a good choice for the absence of thermal instability.

## 5. Conclusion

In conclusion, cavity enhanced frequency doubling at 795 nm is demonstrated with three kinds of nonlinear crystals: PPKTP, LBO and BiBO. The conversion efficiency of each crystal is 58.1% (191 mW - 111 mW), 57.4% (1.34 W - 770 mW) and 35.9% (323 mW - 116 mW) respectively. For the PPKTP crystal, the highest conversion efficiency is obtained, but in order to prevent it from damage, high fundamental power and long-term illumination should be avoided. LBO crystal has good long-term stability, using an anamorphic prism pair or a fiber could be helpful in optimizing the mode of the UV radiation. For the BiBO crystal, the photo-refractive effect of our crystals limit the frequency doubling process, if it could be solved, BiBO crystals could be efficient at the UV regime. In this paper, we investigated and compared the frequency doubling behavior in details. The characters of each crystal that illustrated above could be practical in the fields of nonlinear frequency conversion. In current experiments, the generated UV radiation has satisfied power, good beam quality and stable performance, it is sufficient to be used in the further experiments. To further improve our system, a monolithic cavity design could be established. With all the four mirrors mounted on a whole block, the vibration noise could be well decreased and the better mechanical stability can be expected. For our perspective, the second harmonic of 397.5 nm UV radiation will be used to pump an OPO, the generated squeezed states at 795 nm would be an important source in the precise measurements or the other fields in quantum physics.

**Acknowledgments**

This project is supported by the National Natural Science Foundation of China (61227902, 61475091, and 11274213), and the National Major Scientific Research Program of China (2012CB921601).